\documentstyle[aps,epsfig,float,twocolumn,prl]{revtex}
\newcommand{\be}{\begin{equation}}
\newcommand{\ee}{\end{equation}}

\begin{document}
\twocolumn[\hsize\textwidth\columnwidth\hsize\csname @twocolumnfalse\endcsname
\draft
\title{Breakdown of Conformal Invariance at Strongly Random Critical Points}
\author{M. B. Hastings$^1$ and S. L. Sondhi$^2$}
\address{$^1$ CNLS, MS B258, Los Alamos National Laboratory, Los Alamos,
NM 87545, hastings@cnls.lanl.gov\\
$^2$ Department of Physics, Princeton University,
Princeton, NJ 08544}
\date{October 23, 2000}
\maketitle
\begin{abstract}
We consider the breakdown of conformal and scale invariance in random systems
with strongly random critical points.  Extending previous results on
one-dimensional systems, we provide an example of a three-dimensional system
which has a strongly random critical point.  The average correlation
functions of this system demonstrate a breakdown of conformal invariance,
while the typical correlation functions demonstrate a breakdown of scale
invariance.  The breakdown of conformal invariance is due to the
vanishing of the correlation functions {\it at} the infinite disorder
fixed point, causing the critical correlation functions to be
controlled by a dangerously irrelevant operator describing the approach
to the fixed point.  We relate the computation of average correlation
functions to a problem of persistence in the RG flow.
\end{abstract}
]
\section{Introduction}
The principle of conformal invariance \cite{polyakov} has proven
immensely powerful in the study of continuous phase transitions.
While it is believed to hold at critical points of systems with
short ranged interactions in all dimensions \cite{polchinski}, it
is especially powerful in two dimensions, where it has enabled
a substantial classification of critical theories based on the
representation theory of the Virasoro algebra and its extensions
\cite{senechal}.

Recent interest has focussed on applying this analysis to systems
with quenched randomness in two dimensions; see for
examples Refs.~\cite{bernard,fendley,bthy}.
The formal device that makes this possible is the construction of
translationally invariant field theories whose correlation functions
are the disorder averaged correlators of the random problem, quite
generally via the replica trick and in Gaussian problems, via the
supersymmetry method. (There is at least one notable example where
neither is needed \cite{mudry0}, and conformal invariance follows
straightforwardly.)

Our interest in this paper, is in asking whether the averaged field
theories are {\it necessarily} conformally invariant at their critical
points. We will find that this is not the case in two instances of
what we term strongly random critical points, a category that may well
have considerable overlap with the ``infinite disorder'' critical points 
studied by means of real space renormalization group techniques \cite{infinite}.
Both are localization problems, the first a well studied problem in $d=1$
where conformal invariance is even more powerful than in $d=2$, and the
second is an analogous construction on our part in $d=3$. We also
comment on another well studied member of this family of problems in
$d=2$ which does appear to have a region where conformal invariance holds.
We should note before proceeding further, that while we were motivated by the 
absence of conformal invariance in the one dimensional example, what we
find is a breakdown of scale invariance itself in that the operator product
expansions is anomalous even though two point correlators are algebraic.

Another question appears to be closely connected to the above issues: namely,
how one might recover the {\it distributions} for correlation functions that should
characterize the universal content of random fixed points, from computations
of averages and higher moments. In both of our examples, the typical correlations
are very different from the average ones and are not even algebraic, which suggest
both the difficulty of reconstructing them from the averaged field theory and
why the latter may have to be anomalous. This raises the question of whether the
breakdown of conformal invariance we report may have echoes at other, less
strongly random critical points. We will speculate briefly on this at the end
of the paper.

One rationalization of our results will be that the averaged correlations 
vanish {\it at} the infinite disorder {\it fixed} point, causing the 
behavior of the critical correlation functions to be controlled by a 
dangerously irrelevant operator describing the approach to the fixed point.  
For the typical correlation functions, scale invariance breaks down, while
the average correlation functions have power law behavior at the
level of two point correlators, but break scale invariance for three and
higher point correlations.

Within this interpretation, the power law behavior of two-point
correlation functions is {\it not} the
result of a non-vanishing scaling dimension for the given operators at
the critical point, as in this case one would find scaling for multi-point
correlations as well.  Instead, the power law behavior is
a feature of the leading corrections to scaling.

We begin by reviewing a one-dimensional system exhibiting the given
breakdown in scale and conformal invariance, and show the anomalous
operator product expansion.
Next, we introduce a model system in three
dimensions.  We present evidence that the system is at a critical
point, with relevant perturbations that introduce a localization length
that diverges as the critical point is approached; however, we will
be unable to describe fully the system off-criticality.
We then compute typical correlation functions
in this system and show that they violate scale invariance.  
Using a Liouville field theory to compute average correlation functions,
we show that while the two point functions are power law,
indicative of a critical point and compatible with scale
invariance, multi-point correlation functions violate conformal invariance.

In an Appendix, we note that there exists a SUSY field theory which reproduces
the correlation functions of the Liouville field theory.  Although the
correlation functions are easier to compute in the Liouville theory, the
advantage of the SUSY theory is that it provides a purely local
field theory which exhibits a breakdown of conformal invariance; the
Liouville field theory will involve one global integral which effectively
introduces a long-range interaction.
\section{Breakdown of Conformal Invariance in One Dimension}
The conformal group is the group of transformations which leaves
angles unchanged.  In three, or more, dimensions, the conformal group
is generated by translations, rotations, dilatations, and inversions 
($x^{\mu}\rightarrow \frac{x^{\mu}}{|x|^2}$).  By composing
inversion-translation-inversion, one generates the so-called special conformal
transformations.  In two dimensions, the conformal group is supplemented
by all analytic transformations of the complex plane, while in
one dimension any diffeomorphism is a conformal transformation.

Invariance under translation, dilatation, and rotation fixes the
two point function to be a power law.  Invariance under inversion
fixes the form of the three-point correlation
function.  For example, consider a real scalar field, $\phi$, in
$d>2$ dimensions with action
\be
S=-\frac{1}{2}(\partial_{\mu} \phi)^2.
\ee
Consider the {\it connected} two-point function of $\phi^2$,
$\langle \phi^2(0) \phi^2(x) \rangle\propto |x|^{4-2d}$, a power
law as expected.  The {\it connected} three-point function
is given by
\be
\label{3ex}
\langle \phi^2(0) \phi^2(x) \phi^2(y) \rangle\propto 
|x|^{2-d} |y|^{2-d} |x-y|^{2-d},
\ee
in agreement with inversion symmetry.

In one and two dimensions, conformal invariance imposes
even more stringent requirements on the correlation functions.
In one dimension, conformal invariance requires that,
at a critical point, all correlation functions are constant.
An example of this behavior in statistical mechanics is the one-dimensional
Yang-Lee edge\cite{yl}.

Another example is the following quantum system.  Consider a one-dimensional
Dirac particle with Hamiltonian
\be
\label{dirac}
H=\pmatrix{0 & m(x)+\partial_x \cr m(x)-\partial_x & 0}.
\ee
If we take $m(x)=\overline m$, then, looking at the zero energy Green's
function, the point $\overline m=0$ is a critical point.  In accord with
conformal invariance, the Green's function is a constant.  For non-zero
$\overline m$, the system acquires a correlation length proportional to
$\overline m^{-1}$.

Turning to a random version of this critical point, however,
we find a breakdown of conformal invariance.  
Let $m(x)=\overline m+\delta m(x)$, with $\delta m(x)$ a quenched Gaussian 
random variable with vanishing mean and
\be
\label{gwn}
\overline{\delta m(x) \delta m(x')}=g\delta(x-x'),
\ee
where the overline denotes averaging over disorder.
With this random mass, the Dirac equation is the continuum limit of a 
tight-binding Hamiltonian on
a one-dimensional chain, with randomly chosen hopping elements and
no on-site potential\cite{dyson}.  The zero energy eigenfunctions of this 
problem provide an exception to exponential localization in one-dimension.

For this problem, we will being by summarizing old results on the two-point
functions.  We will note that the two-point function, while invariant
under translation, dilatation, and inversion, is not invariant
under arbitrary diffeomorphisms, thus violating conformal invariance
in one dimension.  Then, we
will show how the correlation functions may be obtained from the
exact zero energy eigenfunctions of the problem, and use this to
compute the three-point function.  This function will not be
invariant under inversion or special conformal transformations, thus
breaking conformal invariance in a way that generalizes to higher
dimensions.

For $\overline m=0$, it has been
shown\cite{dsf,bf,mud1} that the average zero energy Green's function decays
as a power law
\be
\label{pl1}
\overline{G(x,y)}\propto |x-y|^{-3/2}.
\ee
All positive moments of the average Green's function behave similarly.
For convenience, in this paper we choose to focus on a
density of states correlation function.  Let $\psi_a(x)$ be an eigenfunction
of Hamiltonian (\ref{dirac}), with energy $E_a$.  Introduce an infinitesimal
parameter $\eta$.
Define the density of states $\rho(x)$ to be equal to
\be
\rho(x)={\rm Im}(G(x,x))=
{\rm Im}(\sum\limits_a \frac{1}{E_a+i \eta}|\psi_a(x)|^2).
\ee

The two point correlation function of the density of states scales as
Eq. (\ref{pl1}).
On the other hand, if we look at the full probability distribution of the 
density of states at two points, we find that if $\rho(x)$ is of order
unity, then the {\it typical} $\rho(y)$ is of order
\be
\label{se1}
e^{-\sqrt{g|x-y|}}.
\ee

Eq. (\ref{pl1}) is not invariant under general diffeomorphisms,
while Eq. (\ref{se1}) is not scale invariant.  However,
we emphasize that the point $\overline m=0$ is a critical point for
the random system, as for $\overline m\neq 0$,
the average Green's function acquires a diverging correlation length 
proportional to $\overline m^{-2}$, while the typical correlation function
is proportional to $\overline m^{-1}$.  Similarly, away from zero
energy one finds a finite localization length, diverging as
${\rm ln}^2(1/E)$ or ${\rm ln}(1/E)$ for average and typical
correlation functions, respectively.

To derive the above results, consider the
exact zero energy eigenfunctions of the problem:
\be
\pmatrix {e^{\beta(x)} \cr 0}, \; 
\pmatrix {0 \cr e^{-\beta(x)}}
\ee
where $\beta(x)=\int\limits^{x} m(y) \, dy$.
Representing the zero energy Green's function as a resolvent
\be
G(x,y)=\langle x| \frac{1}{H+i \eta} | y \rangle,
\ee
we note that, for a finite system
in the limit $\eta \rightarrow 0$, the Green's function and density
of states are controlled by the zero energy eigenfunction; 
the imaginary part of the
Green's function is equal to $\frac{1}{\eta}\overline\psi(x)\psi(y)$
where $\psi$ is the zero energy eigenfunction.  The zero energy Green's
function is a matrix; the different components of this matrix provide
information on the two different zero energy eigenfunctions.
In the Appendix, a supersymmetric field theory is constructed
such that the zero energy Green's function is a correlation function
of the form
\be
\langle \overline \psi(x) \psi(y) \rangle.
\ee

If the infinite volume limit is taken before the $\eta \rightarrow 0$ limit, 
the resolvent and density of states depend
 on other eigenfunctions beyond the exact zero
energy eigenfunction.  We now show, however, that the scaling properties
of correlation functions for both these quantities
are the same for either order of limits.  Consider eigenfunction
$\psi_a$ at energy $E_a$.  The eigenfunction is a {\it zero} energy eigenfunction
of Hamiltonian
\be
H=\pmatrix{-E & m(x)+\partial_x \cr m(x)-\partial_x & -E}.
\ee
By performing an axial transformation, 
\be
\label{gt}
\tilde \psi_a=
\pmatrix {e^{-\beta(x)} & 0 \cr
0 & e^{\beta(x)}} \psi_a,
\ee
we find that $\tilde \psi_a$ is a zero energy eigenfunction of Hamiltonian
\be
\label{dg}
H=\pmatrix{-E e^{2 \beta(x)}& \partial_x \cr -\partial_x & -E
e^{-2 \beta(x)}}.
\ee
This Hamiltonian has diagonal disorder, and one expects to find
all the eigenfunctions localized; however, for small $E$, the diagonal
terms can be ignored for lengths such that $E e^{2\beta}<<1$.  Since
$\beta$ has rms fluctuations of order $\sqrt{L}$, the diagonal terms
in Hamiltonian (\ref{dg}) can be ignored for $L << {\rm ln}^2(1/E)$,
and the exponential factor from the axial transformation (\ref{gt})
controls the magnitude of $\psi_a$ (we note that the exponential
behavior of $\beta$ ensures that the dominant contribution to
the magnitude of $\psi_a$ arises from
the axial transformation, rather than any fluctuations in 
magnitude of $\tilde \psi_a$).
Thus, we find that over short length scales all the eigenfunctions
track the zero energy eigenfunction in relative magnitude; 
over longer length scales, the non-zero energy eigenfunctions can become 
localized.  

Now consider correlation functions of density of states for finite
$\eta$.  Since the density
of states includes eigenfunctions with energy $E$ of order $\eta$, we
find that $\eta$ sets a length scale in the system (this length is
identical to the average localization length discussed above).
However, on lengths shorter than this scale, the magnitude of
of the eigenfunctions contributing to the density of states can
be obtained from that of the zero energy eigenfunction, up to
constant factors.  Consider a correlation
function of the density of states, $\overline{\prod\limits_{i} \rho(x_i)}$.
We argue that these correlation functions, in the thermodynamic limit with
fixed $\eta$, are equivalent, up to constant factors, to correlation
functions taken in a finite system as $\eta\rightarrow 0$, provided
that in the first case $|x_i-x_j|<<{\rm ln}^2(\frac{1}{\eta})$.
A similar argument may be made for average Green's functions.

Given this equivalence, for the remainder of the paper we consider only
the behavior of the zero-energy eigenfunction.
The log of the exact zero-energy eigenfunction, $\beta(x)$,
executes a random walk, such
that the mean-square fluctuations in the log are proportional to the length
scale $L$.
If we focus on the first of the two eigenfunctions, we see that it is strongly
peaked at a given point, the maximum of $\beta(x)$ (the other eigenfunction
is peaked at the maximum of $-\beta(x)$).  

After normalizing
the eigenfunction, it is exponentially small away from the maximum.
Typically the eigenfunction decays as a stretched 
exponential, giving Eq. (\ref{se1}).  It is possible, though, for
the eigenfunction to have a secondary maximum.  This possibility dominates
the average two-point function.  Eq. (\ref{pl1}) can be obtained
as follows: we wish to compute the probability that $\beta(x)$
has global maxima at $0,x$.   Since the eigenfunction normalization
can be absorbed by a shift in $\beta$, we fix $\beta(0)=\beta(x)=0$.
Everywhere else, $\beta(y)<0$.  Thus, the probability to have maxima
at $0,x$ scales as the return probability of a random walk required to persist
below 0.  This scales as $|x|^{-3/2}$, giving Eq. (\ref{pl1}).
Further, the random walk technique shows that all positive moments have
the same scaling, as any positive moment of the zero energy eigenfunction
is sharply peaked at the maxima of $\beta$.

Tuning away from the critical point by taking $\overline m\neq 0$,
we find that the zero energy eigenfunction is normalizable
only for a finite system, where it describes eigenfunctions decaying into
the system.  The decay of the eigenfunctions is governed by the diverging
correlation length discussed above.

We now turn to the three-point correlation function of the density.
This function can be represented in the supersymmetric field theory as
\be
\langle \overline \psi_1(0) \psi_1(0) \overline \psi_2(x) \psi_2(x)
\overline \psi_3(y) \psi_3(y) \rangle.
\ee
The average three-point correlation function can be computed by
an extension of the random walk argument.  Assume that $0<x<y$.
Then, we consider a random walk that starts at 0, returns to zero at x,
and returns again to 0 at $y$.  This gives the
average three-point correlation function
\be
\label{3p1}
\overline{\rho(0) \rho(x) \rho(y)} \propto |x|^{-3/2} |x-y|^{-3/2}.
\ee
As stated above, Eq. (\ref{3p1}) is not invariant under inversion or 
special conformal transformations.

The three-point function enables us to find the operator product
expansion for the theory.  Given a set of operators, $O_A$, with
dimensions, $d_A$, one expects that
\be
\label{oped}
O_A(0) O_B(x) \rightarrow f_{ABC} x^{-(d_A+d_B-d_C)} O_C.
\ee
Since all positive moments of the density of states scale equivalently,
we need consider only two operators.  One operator is the
identity operator, $I$, while the other operator $O$ is the
density of states operator, with dimension $d$.  
Eq. (\ref{pl1}) suggests $d=-3/4$.  Then, consistency with
the operator product expansion would give
\be
\overline{\rho(0) \rho(x) \rho(y)} \propto |x|^{-3/4} |x-y|^{-3/4} |y|^{-3/4},
\ee
which, like Eq. (\ref{3ex}), obeys inversion symmetry.
However, this is not what we have found.  Thus, we rationalize our
results as follows: the scaling dimension of the operator $O$ vanishes,
as indeed it must in one dimension.  However, the rms fluctuations
in $\beta$ scale as $\sqrt{L}$ and the inverse rms fluctuations
in $\beta$ tend to zero as $L\rightarrow \infty$;
equivalently, the statistical properties of the zero energy eigenfunction
are unchanged under scaling $x\rightarrow x/L,g \rightarrow L g$.
If, however, we take $g$ divergent,
we find that the zero-energy eigenfunction is localized at a point and
all correlation functions vanish.  Thus, $g^{-1}$ is
a dangerously irrelevant variable.  In the scaling
limit, defined as the limit in which the separation between points
in correlation functions tends to infinity, all correlation functions are 
controlled by the vanishing of this variable as the fixed
point is approached.  Similarly, $f_{ABC}$ vanishes as the fluctuations
in $\beta$ diverge.  However, what we have
found above is that $f$ {\it vanishes as a power law as the fixed
point is approached}.  Thus, the power law (\ref{pl1}) is {\it not}
the result of $d \neq 0$, but the result of $f\rightarrow 0$ as the
fixed point is approached.

We note that the behavior we have found is similar to that found
using a real-space RG approach\cite{dsf} on the discrete tight-binding
version of this model.  In that approach, one renormalizes
the distribution of hopping elements, reducing a cutoff on the strength
of the hopping.  In the scaling
limit, almost all the hopping elements become vanishingly small compared
to the cutoff; the fraction of hopping elements (vanishing as
a power law in the scaling limit) that are of order the cutoff control
the correlation functions.
\section{A Model System in Three Dimensions}
To demonstrate the generality of this breakdown of conformal invariance,
we will look for higher dimensional examples.  Using a real-space 
renormalization group, some examples have been found of infinite disorder 
quantum critical points two or more dimensions\cite{2dis}.  However, these
are {\it interacting} quantum systems and can only be described
by 2+1 or 3+1 dimensional theories with a privileged time coordinate, 
so they would not be expected to have conformal invariance, although
the question of scale invariance in these systems remains an interesting
problem for future work.  There is also some evidence\cite{2hop} 
that a two-dimensional random hopping model could have an infinite disorder
critical point.  However, there is as yet no evidence for breakdown of 
conformal invariance in this model.  Therefore, we will construct a model 
system in three-dimensions that provides a strongly random critical point 
without conformal invariance.  

After presenting the Hamiltonian and eigenfunctions, we will present evidence
that the system is indeed at a critical point between two different
localized phases.  Further evidence for the criticality of our model
will result from behavior of typical and average correlation functions
that we find below.

Consider the following Hamiltonian
\be
\label{h3d}
H=\pmatrix{0 & (\partial_{\mu}-A_{\mu})^2 \cr (\partial_{\mu}+A_{\mu})^2 & 0},
\ee
where 
\be
A_{\mu}(x)=\partial_{\mu} \beta(x).
\ee

The Hamiltonian has two exact zero energy eigenfunctions.  They are
\be
\label{z3}
\pmatrix {e^{\beta(x)} \cr 0} , \;
\pmatrix {0 \cr e^{-\beta(x)}}.
\ee

We pick $\beta(x)$ with probability distribution $e^{-S[\beta]}$ where
\be
\label{sbeta}
S[\beta]=\frac{1}{2g}\int d^3 x\, (\partial^2 \beta)^2.
\ee

We pick this distribution of disorder as the best balance between having
too little and too much disorder, while retaining a local distribution
of disorder.  Let us show that for less disorder, the critical point
would not be strongly random, while for more disorder we might lose the 
critical point completely.

With this distribution for $\beta$, the mean-square fluctuations in
$\beta$ are proportional to $L$, the length scale of the system, just as
in the one-dimensional system the fluctuations in $\int m(y) \, dy$ were
proportional to the length of the system.  This increase
in fluctuations with length scale enables us to reach a strongly
random fixed point.  

Suppose instead that we had chosen a disorder distribution with smaller 
fluctuations in $\beta$.  If we had chosen a two-dimensional
system with $S[\beta]=\int d^2 x \, (\partial_{\mu} \beta)^2$, the fluctuations
in $\beta$ would only grow logarithmically with $L$, the exponential
of $\beta$ would behave as a power law, and we find
a conformally invariant fixed point\cite{mud2}.  If we had considered a
three-dimensional system with $S[\beta]=\int d^3 x\, (\partial_{\mu} \beta)^2$,
we would have found a fixed point with vanishing disorder.

On the other hand, for the given distribution of disorder, we can
make an argument that the system is at a critical point.
For the given distribution of disorder,
by letting $A_{\mu}=\partial_{\mu}\beta(x)+J_{\mu}$, for $J_{\mu}\neq 0$,
we can tune away from criticality.  As in the one-dimensional example,
we find that the eigenfunction
is normalizable only for a finite system and describes eigenfunctions decaying 
into the system.  As now the zero energy eigenfunctions are
\be
\pmatrix {e^{\beta(x)+J_{\mu}x^{\mu}} \cr 0} , \;
\pmatrix {0 \cr e^{-\beta(x)-J_{\mu}x^{\mu}}},
\ee 
the typical eigenfunction decays with a length proportional to $|J|^{-1}$.  
This demonstrates that there is a diverging length scale associated
with the given problem, implying that we are at a critical point.
If we consider a system in a square box, with $J_\mu=(j,0,0)$, as
we tune $j$ from positive to negative there is a phase transition between
eigenfunctions localized on opposite faces of the box.
It should also be possible to tune away from criticality by considering
Green's functions at non-zero energy.  This is a more difficult task, for
future work.

However, if we had considered a two-dimensional system with
$S[\beta]=\int d^2 x\, (\partial^2 \beta)^2$, the mean-square
fluctuations in $\beta$ would grow as $L^2$.  We would
find that the typical correlation functions for eigenfunction (\ref{z3})
would decay exponentially.  In the case of the system in a box,
even for vanishing $J$ there would be eigenfunctions exponentially localized
on the faces of the box, and the phase transition would
occur at a non-vanishing value of $J$.  Therefore, for this two-dimensional
distribution we would not have such strong evidence that the system is
critical.

Finally, let us emphasize that the distribution of disorder we
have chosen in Eq. (\ref{sbeta}) is {\it local}.
As a result, the SUSY field theory in the Appendix is also local, and
so the breakdown of conformal invariance is not a result of long-range
interactions.
\section{Typical Correlation Functions and One-Point Functions}
Let us focus on the first zero energy eigenfunction of Eq. (\ref{z3}) and
write $\psi(x)=e^{\beta(x)}$.
We normalize $\psi$ such that $\int d^3 x\, \psi^2(x)=1$.  
In exact analogy with the one-dimensional system, we find that if
$\psi(x)$ is of order unity, then
\be
\label{se3}
\psi_{\rm typ}(y)\propto e^{-\sqrt{g|x-y|}}.
\ee

We can also consider a typical one-point function, the magnitude
$|\psi(x)|^2$.  Of course, on average, this function is equal to
$\frac{1}{L^3}$.  Typically, however, it is of order 
\be
\label{1p3}
e^{-\sqrt{gL}}.
\ee

Clearly, Eqs. (\ref{se3},\ref{1p3}) violate scale invariance.  We will
next turn to the average correlation functions.  They will not violate
scale invariance, but they will violate conformal invariance.  The
advantage of considering average correlation functions is that, as shown
in the Appendix, they can be obtained as correlation functions in a
purely local field theory.
\section{Multifractals and Liouville Field Theory}
Let us look at average correlation functions of the form
\be
W_q(x,L)=\overline{|\psi(x)\psi(0)|^q},
\ee
where the overline denotes ensemble averaging.  From the
scaling behavior of $W_q$, we can compute the scaling of
$\overline {\rho(x)^{q/2}\rho(0)^{q/2}}$.

The inverse participation ratio, 
$W_q(L)=W_q(0,L)$, is expected to vary as a power
law
\be
W_q(L)\propto \frac{1}{L^{3+\tau(q)}},
\ee
while the two-point function is expected to vary as
\be
\label{mf}
W_q(x,L)\propto \frac{1}{L^{3+\tau(q)}}\frac{1}{x^{\sigma(q)}}.
\ee
For localized eigenfunctions, $\tau(q)=0$, while for a plane wave eigenfunction
$\tau(q)=3(q-1)$.  In some localization problems, one finds multifractal scaling
for the $\tau(q)$ at criticality; see ref.~\cite{jans} for a review.

The typical correlation functions of the previous section can be considered
as exponentials of ensemble averages of logarithms of correlation functions.
They can be obtained from $q \rightarrow 0$ limits of the average correlation
functions.

We will now introduce a Liouville field theory\cite{tsv,bf} to
compute the functions $W_q$.  The correct normalization of eigenfunction
$\psi$ can be obtained by
\be
\psi(x)=N^{-1/2} e^{\beta(x)},
\ee
where
\be
N=\int d^3 x \, e^{2\beta(x)}.
\ee
The correlation functions $W_q$ can be written as
\be
W_q(x,L)=\frac{1}{Z_0} \int [d\beta] N^{-q} e^{q\beta(0)}e^{q\beta(x)}
e^{-S[\beta]},
\ee
where $Z_0=\int [d\beta] e^{-S[\beta]}$.  The normalization can be
exponentiated by
\be
N^{-q}=\frac{1}{\Gamma(q)}\int\limits_{0}^{\infty} d\omega
\omega^{q-1} e^{-\omega N}.
\ee

As a result,
\be
\label{lv}
W_q(x,L)=\frac{1}{Z_0 \Gamma(q)}\int\limits_{0}^{\infty} d\omega [d\beta]
\omega^{q-1} e^{q\beta(0)} e^{q \beta(x)} e^{-S[\beta]-w e^{2 \beta}}.
\ee
This provides a three-dimensional Liouville field theory, with the
strange feature that the action involves fourth derivative terms instead
of second derivative terms.
\section{Average Correlation Functions}
We will consider the average correlation functions.  We will demonstrate
that the exponents $\tau(q)$ and $\sigma(q)$ are independent of $q$.  We will
be unable to calculate $\sigma(q)$ exactly, but we will argue that the power law
behavior of Eq. (\ref{mf}) is obeyed for a finite $\sigma(q)$, leading
to conformal invariance for the two point function.   The behavior of
the average correlation function is again related to problems of
persistence in nonequilibrium systems.  Then, as in one dimension, 
we will show using OPEs of the theory that the three point function
breaks conformal invariance.

If we look at the scaling behavior of Eq. (\ref{lv}), we find that as the 
length scale $L$ increases, $\beta$ increases proportional to
$L^{1/2}$, and so $e^{q \beta}$ becomes sharper.  Effectively, $q$
increases proportional to $L^{1/2}$, and so for any $q>0$, the
scaling behavior of the correlation functions is the same, so that
$\tau(q)$ and $\sigma(q)$ are independent of $q$.  The $q\rightarrow 0$ and
$L\rightarrow \infty$ limits do not commute.

In the large $q$ limit, and therefore in the scaling limit, the two-point 
function is dominated by field configurations in which $\beta$ reaches its 
maximum at $0$ and $x$.  The one-point function is dominated by configurations
in which $\beta$ reaches its maximum at $0$; the probability of this happening
is proportional to $1/L^3$ and so $\tau(q)=0$, characteristic of a localized
eigenfunction.

To find $\sigma(q)$, we need to look at the scaling of the following 
functional integral
\be
\label{wall}
\int [d\beta] e^{-S[\beta]-f(\beta)} \rho_1[\beta]
\rho_2[\beta].
\ee
Here, $\rho_1[\beta]=\delta(\beta(0))$ and
$\rho_2[\beta]=\delta(\beta(x))$, while
$f(\beta)=0$ for $\beta\leq 0$ and $f(\beta)=\infty$ for $\beta>0$,
so that the potential $f$ has a hard wall\cite{bf} at $\beta=0$.
This reproduces the desired constraint that $\beta$ have a maximum at
both $0$ and $x$.
Let us
define an operator $O(x)$ that inserts a factor of $\delta(\beta(x))$
in Eq. (\ref{wall}).  Then $W_q(x,L)=\langle O(0) O(x) \rangle$.

We will proceed with a renormalization group technique as follows:
initially the functionals $\rho_1,\rho_2$ define $\delta$-function distributions
for $\beta$ at $0,x$.  We will replace these with probability distribution
functionals for $\beta$, and for the normal derivative of $\beta$,
 on the surface of a sphere at radius $a$ around points
$0,x$.  Initially $a$ will be small, but we will integrate out the
field $\beta$ in a shell between $a$ and $a+{\rm d}a$, and compute the change
in the probability distribution as $a$ is increased.
In one-dimension, this amounts to computing the probability distribution
of a random walk, persisting below zero, at points $0\pm a,x\pm a$.

As the RG above is carried out with spheres centered on $x,0$,
at $a=|x/2|$ the two spheres merge, and one must match boundary conditions.
On the surface of the merged spheres, the field $\beta$ is of order 
$\sqrt{|x|}$.  To compute the scaling of the two point function, one must 
compute the probability that $\beta$ persists below zero on the surface of both
spheres up to scale $|x/2|$, with matched boundary conditions
(we will discuss this computation below).
After the spheres merge, one has a single surface propagating outwards
as $a$ is increased, but the persistence probability on this surface
is independent of $x$.

Given the two-point function, consider the three-point correlation function
$\langle O(0) O(x) O(y) \rangle$, with $|x|<<|y|$.
At the scale $|y|$, the field $\beta$ is of order 
$\sqrt{|y|}>>\sqrt{|x|}$.
Therefore, for the purpose of computing the scaling of the
three-point function with respect to $y$, we can simply assume that $\beta$
is pinned at the value zero near the point $0$.  Thus, the
two operators $O(0),O(x)$ fuse into a single operator $O$.  
\be
\label{ope}
\lim\limits_{x\rightarrow 0} O(x) O(0)=x^{-\sigma} O(0).
\ee
The exponent as $\sigma$ in the above equation 
is exactly $\sigma(q)$ for $q>0$, as
the two-point function can be computed by fusing the two
operators $O$ into a single operator, yielding a 
one-point function, for which
\be
\label{1pt}
\langle O(0)\rangle=1/L^3.
\ee
Therefore,
\be
\label{2p3}
\langle O(0) O(x) \rangle \propto x^{-\sigma}.
\ee

We now discuss the procedure for computing $\sigma$.  First we
consider the problem of persistence in $\beta$ for a single
sphere, and then show how to
obtain $\sigma$ from the persistence problem by matching boundaries.  However, 
the problem in persistence is sufficiently difficult that we
will not attempt to obtain a numerical value for $\sigma$.
The related persistence problem will at least enable us to
argue that $\sigma$ is finite so that correlation functions
are power law.

Decomposing $\beta$ into spherical harmonics,
\be
\beta(a,\theta,\phi)=\sum\limits_{l,m}f_{lm}(a)Y_{lm}(\theta,\phi),
\ee
the probability distribution functionals can be written as
\be
\rho[f_{lm}(a),\partial_a f_{lm}(a)].
\ee

First, let us consider the probability that, for a single sphere,
$\beta$ does not cross the hard wall up to scale $a$.
Defining
\be
g_{lm}(T)=e^{-T/2} f_{lm}(e^{T}),
\ee
we can write the action $S[\beta]$ as
\be
\label{sc}
\sum_{lm} \int \frac{1}{2}
\Bigl(3/4+l(l+1)+2\partial_T+\partial^2_T)g_{lm}(T)\Bigr)^2
{\rm d}T.
\ee

With this scaling of field, and logarithmic transform of the RG scale,
the problem becomes that of persistence in a Gaussian
stationary process\cite{pmath}.  
With action given by Eq. (\ref{sc}), the correlation function
\be
\langle g_{lm}(T) g_{lm}(T') \rangle
\ee
is exponentially decaying in $|T-T'|$.  Therefore, for a finite number of fields
$g$, the persistence probability is exponentially decaying in $T$,
and decays as $a^{-\theta}$ for some power $\theta$.  We
have an infinite number of fields $g_{lm}$, but for large $l$ they
have very small fluctuations, so it seems likely that $\theta$ tends to
a finite limit as we include fields with greater and greater $l$.

In principle it is possible to compute $\theta$ for a finite number of
$g$.  However, since the process is non-Markovian due to the fourth
derivatives it is quite difficult; even for the simpler problem
of one field, with an action containing only fourth derivative and no other 
terms, this problem\cite{diff} was solved only relatively recently\cite{d2}.

For a finite system, without periodic boundary conditions, the
persistence exponent determines the probability that the maximum of
the eigenfunction will be located at a given interior point, instead of
at the boundary.

Given the persistence exponent, we can compute $\sigma$.  At
scale $a=|x/2|$, the spheres start to merge.  Beyond this scale,
we have only one surface.  Immediately before
the spheres touch, the probability that
$\beta$ is a maximum at $0$ in the sphere centered around $0$ is 
\be
\label{1ph}
|x|^{-\theta}.
\ee
The probability that $\beta$ is a maximum at $x$
in the sphere centered at $x$ is the same. 
The product is $|x|^{-2\theta}$.  

However, at the point where the two spheres meet, the values of 
$\beta$ and $\partial \beta$
must be equal on both spheres.  This is a matching of boundary
conditions.  To match boundary conditions, the scaled variables $g$
must be equal on both spheres.  Since $g$ is stationary, the probability
that the scaled variables are equal is independent of the RG scale $T$.
However, for $\beta$ to be equal on both spheres, there is an additional 
Jacobian which gives rise to a further power law correction.  In
this case, the mean square
fluctuations in $\beta$ grow as $\sqrt{L}$, while the fluctuations
in $\partial \beta$ do not grow in $L$.  Therefore, the probability that 
$\beta$ is a maximum at both $0$ and $x$, and that the boundary conditions
match (equivalently, that the maxima at $0$ and $x$ coincide) is
$|x|^{-2\theta-1/2}$.  

This is to be compared to Eq. (\ref{1ph}) for
only one sphere, so that the relative probability to find two
maxima is
\be
\sigma=2\theta+1/2-\theta=\theta+1/2.
\ee
We see this explicitly in the problem of the random Dirac equation
(\ref{dirac}) in one-dimension, where the persistence exponent
$\theta=1$ (typically in the literature, this is quoted as $\theta=1/2$
as one typically computes only the persistence probability for a
nonequilibrium process going forward in time; our spheres move outwards
in both directions and so the exponent is doubled).
The average correlation function decays as $x^{-3/2}=x^{-\theta-1/2}$.

Even without a precise value for $\sigma$, we can still demonstrate
the breakdown of conformal invariance.
The two-point function calculated above exhibits scale and conformal
invariance in 3 dimensions.  However, as in one dimension,
the operator product expansion 
is not consistent with scaling.
From the two-point function, one would expect that $O$ would have scaling
dimension $\sigma/2$.  However, Eq. (\ref{ope}) gives
a scaling dimension of $\sigma$ for $O$.  The trouble is that 
Eqs. (\ref{ope},\ref{2p3}) are not simultaneously consistent with
scaling.

To make the contradiction more concrete, consider the three point function
$W_q(x,y,L)=\overline{|\psi(0)\psi(x)\psi(y)|^q}=\langle O(0) O(x) O(y)\rangle$
in the limit when $|x|<<|y|$.
Although we were unable to compute $\sigma$, we are able to use the
operator product expansion to
demonstrate a breakdown of conformal invariance.  

In the limit $|x|<<|y|$, the operator product expansion gives
\be
\label{3pt}
W_q(x,y,L)\propto L^{-3} |x|^{-\sigma} |y|^{-\sigma}.
\ee
This is similar to Eq. (\ref{3p1}) found in one dimension.
Eq. (\ref{3pt}) violates
inversion symmetry with respect to a point $|z|$ where $|z|<<|x|$.

As in one-dimension, we interpret this to mean that
$O$ has {\it vanishing} scaling dimension, but that
all the correlation functions of $O$ vanish at the infinite disorder fixed 
point.  As the theory approaches the fixed point, the
average correlation functions vanish as a power law, so that if
one only considers the two-point functions one obtains the wrong scaling
dimension for $O$.  The inverse of the mean-square fluctuations in $\beta$
is a dangerously irrelevant operator that describes
the approach to the fixed point; when this operator vanishes, so do
the correlators.  
\section{Conclusion}
We have considered the breakdown of conformal invariance in certain
strongly random fixed points.  The breakdown of conformal invariance is
related to a power law vanishing of correlation functions approaching
the fixed point.

Although we were unable to compute the exponent $\sigma$ precisely, we
were able to obtain some understanding of the operator product
expansion, and to relate
$\sigma$ to a persistence exponent in a nonequilibrium process.
The connection to persistence is not surprising; the RG flow describing
the approach to the fixed point is non-equilibrium, as the flow approaches
the fixed point but never reaches it.  In contrast to ordinary critical
points, it is the approach to the fixed point, rather than the fixed point
itself, that controls the exponents.

Within other strong disorder RGs, such as for the transverse field Ising
model\cite{dsf,2dis}, average correlation functions are controlled by
the probability that a given pair of sites will remain within the ordered phase
as the RG is run up to the scale of the separation between the sites.  This
is also a question of persistence.

Interestingly, for other random fixed points, such as the
problem of two-dimensional Dirac fermions in random magnetic field\cite{mud2}
or the problem of Anderson localization\cite{akl}, it has been found that
averages of low moments of correlation functions exhibit conformal
invariance, but for sufficiently high moments conformal invariance
breaks down.  Shapiro showed for the Anderson localization problem\cite{shapiro}
that the breakdown of conformal invariance could be described by
a universal distribution of conductance with a broad tail, such
that higher moments were divergent when evaluated with the
fixed point distribution of conductance.  As a result, the higher
moments are controlled by the approach to the universal distribution.

We would like to close with a conjecture.  For the strongly random
systems we consider, all positive moments of the correlation functions
have the same scaling behavior.  For systems with weaker randomness, one
observes different scaling dimensions for the various moments, $q$.  However,
there is an expectation that the scaling dimension will saturate, and
become independent of $q$ for sufficiently large $q$.  At this point,
one may calculate correlation functions with $q=\infty$.  Consider
for definiteness a disordered Ising model. In
this case, the only contribution to an $n$-point correlation function
of spins will arise when all $n$ points are in a locally ferromagnetic
region.  In this case, we conjecture that the computation of correlation
functions will again become a problem of persistence in the RG flow,
that the system will stay within the ferromagnetic phase.  Therefore,
we conjecture that a breakdown of conformal invariance will be seen
for sufficiently large moments in other systems as well.
\section{Appendix: A SUSY Field Theory}
In this Appendix, we will show that there exists a supersymmetric
field theory\cite{efetov} which reproduces the average correlation
functions computed above in the Liouville field theory.  Variants of the
supersymmetric field theory can be used to describe the one-dimensional
and three-dimensional problems considered.  The construction
of the supersymmetric field theory will show that there exists a purely
local field theory with a finite number of fields which violates
conformal invariance at the critical point.  
In the Liouville theory,
the integral over $\omega$ in Eq. (\ref{lv})
introduces a non-local interaction, which might otherwise appear to be
responsible for the breakdown of conformal invariance.

We are interested in the resolvent
\be
G(x,y)=\langle x| \frac{1}{H+i \eta} | y \rangle.
\ee
If the $\eta\rightarrow 0$ limit is taken before the infinite
volume limit, the resolvent is determined by the exact zero-energy
eigenfunction.  As discussed above, all the localization properties
near zero energy are controlled by the zero energy eigenfunction.

The computation of averages of $G$ follows standard techniques\cite{efetov}.
Introducing a superfield $\Phi$, with bosonic component
$\phi$ and fermionic component $\psi$, the resolvent is
\be
\frac{1}{Z}\int [d\beta] e^{-S[\beta]} e^{-\overline \Phi(i H-\eta) \Phi}
\psi(x) \psi(y),
\ee
where
\be
Z=\int [d\beta] e^{-S[\beta]}.
\ee

By introducing a set of $m$ superfields $\Phi_i$, we can compute $m$-th
moments of the correlation functions.  Let us emphasize that
we can take $m$ to be fixed, but large, and then we have a single
field theory from which high moments and multi-point correlation
functions can be computed.

In order to compute the typical correlation functions of section IV,
we would need to introduce an infinite set of superfields to compute
all moments, and then obtain the distribution of correlation functions
from the moments.  Alternately, we could compute them from the
Liouville field theory as a limit of the $q$-th moment of the correlation
functions as $q \rightarrow 0$.  In any event, we do not know how to obtain
them from a purely local field theory.

In the SUSY field theory, $\beta$ does not acquire any dynamics as $\Phi$
is integrated out.  Physically, this is the statement that $\beta$ is
quenched.  If $\beta$ were not quenched, one would expect $\beta$ to
acquire a second derivative term under the RG, and then the theory would
flow to a fixed point with finite fluctuations in $\beta$.  The fact
that disorder is quenched is important for the existence of a strong
disorder critical point.
\section{Acknowledgements}
We would like to thank John Cardy and the Aspen Center for
Physics for a conversation (with SLS) that led to this work.
We would also like to thank K. Damle and O. Motrunich for sharing their
unpublished results on the two dimensional hopping model,  
D. Huse and E. Fradkin for useful discussions, and C. Mudry for 
extremely helpful comments on an earlier draft.
We would like to  acknowledge support from DOE grant W-7405-ENG-36
(MBH) and NSF grant No. DMR-9978074 and the David
and Lucille Packard Foundation (SLS).

\end{document}